# Unified Bayesian estimator of EEG reference at infinity: rREST

Shiang Hu[1], Dezhong Yao[1], Pedro A. Valdes-Sosa[1,2]*

[1]The Clinical Hospital of Chengdu Brain Science Institute, MOE Key Lab for NeuroInformation, University of Electronic Science and Technology of China, Chengdu, China
[2]Cuban Neuroscience Center, Havana, Cuba
Corresponding Author: Pedro A. Valdes-Sosa

Email: pedro.valdes@neuroinformatics-collaboratory.org

## Abstract

The choice of reference for the electroencephalogram (EEG) is a long-lasting unsolved issue resulting in inconsistent usages and endless debates. Currently, both the average reference (AR) and the reference electrode standardization technique (REST) are two primary, apparently irreconcilable contenders. We propose a theoretical framework to resolve this reference issue by formulating both a) estimation of potentials at infinity, and, b) determination of the reference, as a unified Bayesian linear inverse problem, which can be solved by penalized maximum likelihood inference. We find that AR and REST are very particular cases of this unified framework: AR results from biophysically non-informative prior; while REST utilizes the prior based on the EEG generative model. To allow for simultaneous denoising and reference estimation, we develop the regularized versions of AR and REST, named rAR and rREST, respectively. Both depend on a regularization parameter that is the noise to signal variance ratio. Traditional and new estimators are evaluated with this framework, by both simulations and analysis of real resting EEGs. Towards this end, we leverage the MRI and EEG data from 89 subjects which participated in the Cuban Human Brain Mapping Project. Generated artificial EEGs—with a known ground truth, show that relative error in estimating the EEG potentials at infinity is lowest for rREST. It also reveals that realistic volume conductor models improve the performances of REST and rREST. Importantly, for practical applications, it is shown that an average lead field gives the results comparable to the individual lead field. Finally, it is shown that the selection of the regularization parameter with Generalized Cross-Validation (GCV) is close to the 'oracle' choice based on the ground truth. When evaluated with the real 89 resting state EEGs, rREST consistently yields the lowest GCV. This study provides a novel perspective to the EEG reference problem by means of a unified inverse solution framework. It may allow additional principled theoretical formulations and numerical evaluation of performance.

## 1. Introduction

The human electroencephalogram (EEG) has been an indispensable technology for both cognitive and clinical neuroscience for almost 90 years. Ultrahigh temporal resolution, low cost, and noninvasiveness single it out as a translational tool of choice to study the brain. Nevertheless, two main drawbacks of EEG detract from its ability to localize the brain activity: i) spatial blurring due to volume conduction, ii) the inherent indeterminacy of potentials measurements which are always carried out with respect to a given reference (Teplan, 2002). Spatial blurring is being addressed by advanced source imaging techniques that however will not be the focus of our attention. We will rather concentrate on the vexing and still incompletely resolved 'EEG reference problem'. To precisely define this issue, we note that it is due to the intrinsic nature of EEG recordings that are the measurement of potential differences between two sites shown in **Figure 1**. Ideally, one would like to record the potentials of an 'active electrode' that is only picking up the activities due to a few brain structures in comparison to a neutral 'reference electrode' with zero activity. One might think that such a reference electrode could be placed at infinity, yielding the ideal potentials $\varphi_\infty$. However, this would not work in practice, since this configuration would serve as an antenna, picking unwanted activity from the environment. Some researchers therefore experimented with reference electrode placed on the body so that EEG differential amplimers could eliminate environmental noise with high common mode rejection ratio. Unfortunately, because there is no neutral or inactive point upon the body, these proposals are also inadequate. A physical neutral reference seems therefore out of our reach.

However, the non-neutrality of the reference has consequences cascading through the following processing stages, including the final statistical result. In view of the failures of physical references, attention was turned to the construction of 'virtual' estimators of the neutral references, namely, virtual estimators of $\varphi_\infty$.

One popular virtual estimator is the 'average reference' (AR, **Figure 1(d)**), which based on the following logic: i) the integral of the electrical potential over a sphere, due to a current source inside it, is zero (Goldman, 1950; Offner, 1950), ii) the head can be approximated as a sphere. iii) therefore, a neutral reference may be obtained by summing or averaging the activities of all electrodes. Re-referencing proceeds by subtracting this average from all channels. Unfortunately, recent work (Yao, 2017) has shaken the theoretical foundation of AR: the potential integral for a realistic head surface is not zero.

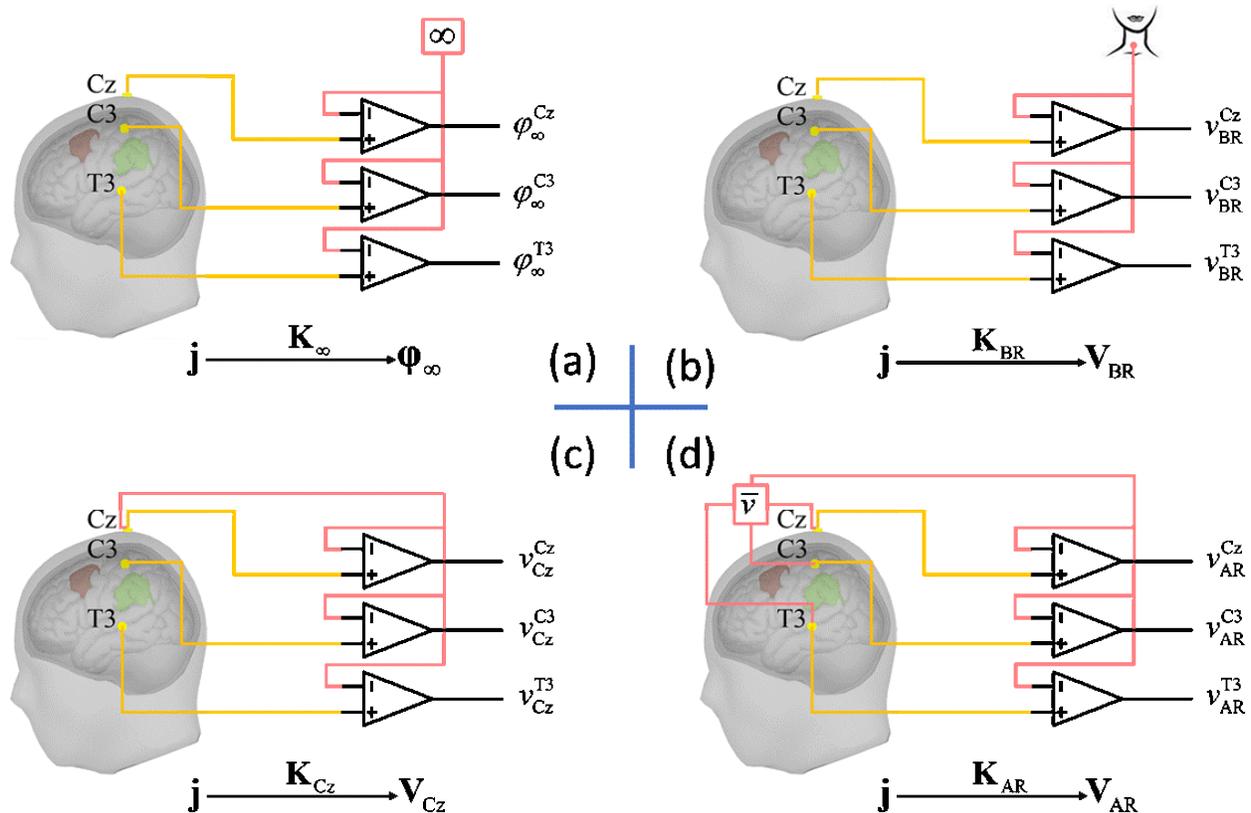

**Figure 1, EEG reference problem.** EEG recordings measure the potential differences between active electrodes marked in yellow (only 3 for illustration) and reference marked in red which may be **(a)** a point at infinity, **(b)** an electrode placed on the body, e.g. the base of neck, **(c)** the cephalic reference (here, Cz), **(d)** average reference --- the mean of potentials over all active electrodes. All reference techniques **(b)-(d)** attempt to approximate infinity reference in **(a)**. The issue is that the all proposals results in different EEG waveforms originating the 'EEG reference

problem'. It should be noted that with the identical source activities $\mathbf{j}$, the different waveforms ($\boldsymbol{\varphi}_\infty$, $\mathbf{V}_{PBR}$, $\mathbf{V}_{Cz}$, $\mathbf{V}_{AR}$) could be taken as well as the outcome of different lead fields ($\mathbf{K}_\infty$, $\mathbf{K}_{PBR}$, $\mathbf{K}_{Cz}$, $\mathbf{K}_{AR}$).

A more biophysically-based virtual estimator of $\boldsymbol{\varphi}_\infty$ can be obtained by the reference electrode standardization technique (REST, **Figure 2**) which directly estimates the ideal potentials referenced to a point at infinity (Yao, 2001). REST uses a head model and equivalent sources to localize source activities, then project the source activities to electrodes — now with reference to infinity. Early work on REST was based upon a simple spherical head model. It was soon shown that EEG power maps (Yao et al., 2005), ERP peak values and latencies (Li and Yao, 2007) did, in fact, critically depend on the choice of reference. In a further study, it was shown (Tian and Yao, 2013) that scalp statistical parametric mapping with REST for audiovisual stimulus evoked potentials provided closer correspondence to the source localization by low resolution electromagnetic tomography than that with AR. These encouraging results about REST have been bolstered by several simulation experiments. Using a spherical head model for simulation, (Marzetti et al., 2007; Qin et al., 2010) indicated that REST led to better estimates of EEG spectra and coherence than AR. Several papers unsurprisingly showed that realistic head model for REST gives superior results for the reconstruction of simulated EEG scalp topographies (Liu et al., 2015), functional connectivity (Chella et al., 2016) and bispectral analysis (Chella et al., 2017).

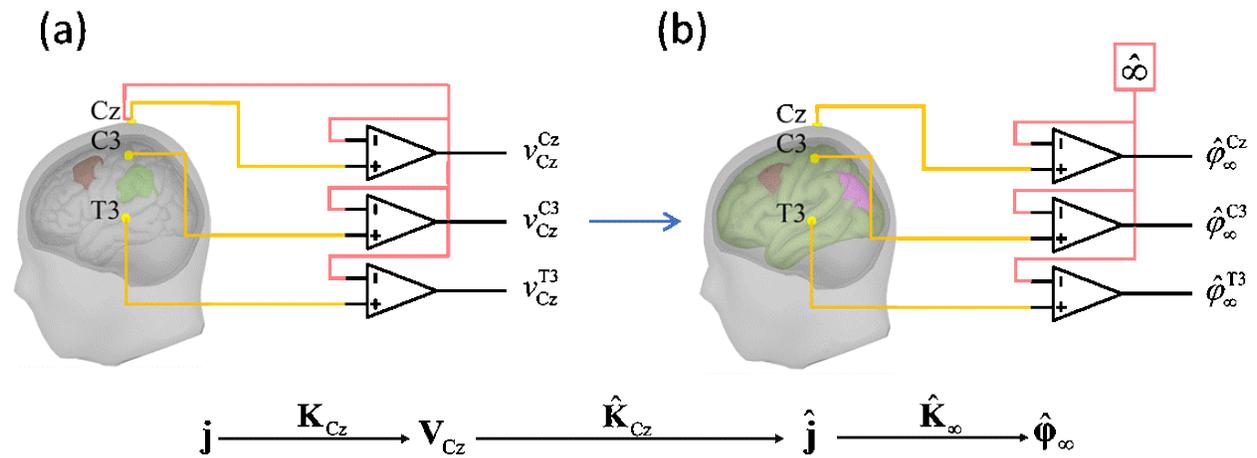

**Figure 2, Diagram of REST. (a)** measured EEG potentials $\mathbf{V}_{Cz}$ with reference to Cz; $\mathbf{j}$ and $\mathbf{K}_{Cz}$ are the actual source activities and the actual lead field with reference to Cz, respectively **(b)** for REST, one first builds the lead field $\hat{\mathbf{K}}_\infty$ with estimated head model and equivalent sources, then transforms the lead field with reference to Cz as $\hat{\mathbf{K}}_{Cz}$; with this lead field, one enables to estimate the equivalent source activities $\hat{\mathbf{j}}$; after, the equivalent source activities are taken the forward calculation through $\hat{\mathbf{K}}_\infty$ to $\hat{\boldsymbol{\varphi}}_\infty$ which is the approximation of EEG potentials at infinity.

In spite of these suggestive results in favor of REST, there is still an intense and to a certain extent unresolved debate on which reference is preferable (Kulaichev, 2016; Nunez, 2010b). The lack of resolution is due to that simulation studies, while useful, are not enough to demonstrate the superiority of one reference technique over another. In addition to simulations, the evidence is needed on which reference achieves the 'best fit' to actual data'. The choice of the best model is a well-studied problem in modern statistics (Christian P. Robert, 2007) and can be resolved by model selection criteria that approximates the Bayesian model evidence (Konishi and Kitagawa, 2008). However, to apply these techniques, an explicit Bayesian model of the 'EEG reference problem' must be stated. Thus, one of the primal goals for this study is to uncover a unified estimator of EEG reference at infinity.

In the current study, we formulate, to our best knowledge for the first time, the 'EEG reference problem' as a generalized Bayesian inverse problem. One surprising consequence of this approach is the insight that AR and REST share the same model and just differ in the prior distribution for the covariance of EEG potentials at infinity. On the one hand, assuming uncorrelated activities over electrodes leads to the AR estimator. On the other hand, if the correlations between electrodes are assumed to be caused by sources filtered through a volume conductor model, the resulting estimator is REST.

Our theoretical formulation will allow us to examine different reference estimators within a common statistical framework. We note that the REST estimator (Yao, 2001) was originally defined for high S/N ratios. In some situations, this might be unrealistic, since the scalp EEG may have quite low signal to noise ratio (Bigdely-Shamlo et al., 2015; Ferree et al., 2001; Guruvareddy, 2013; Lemm et al., 2006). Our framework allows using regularization technique as a way to accommodate noise in the data (Phillips et al., 2002, 2005). The regularized version of REST is developed which we all 'rREST'. It is evident that a regularized version of AR is also possible, which we term as 'rAR'. AR and REST are just the special cases of rAR and rREST when regularization parameter tends to zero, respectively.

We further investigate the effect of the volume conduction model on rREST. To avoid the 'inverse crime' (Kaipio and Somersalo, 2007), the volume conduction model used in the simulation should be different with the one used to generate EEG potentials. We call this 'volume conduction model matching problem' for REST that may produce the spurious results in simulation. Although equivalent source models are used for REST in simulation, the volume conduction model matching problem still cannot be neglected (Hu et al., 2017). Within this framework and using extensive simulations, the performances of AR, REST, rAR and rREST are compared in terms of the relative error of estimation of $\boldsymbol{\varphi}_\infty$. Additionally, in these simulations, we explore the performances of the model selection criteria for selecting the regularization parameter.

Finally, we assess their performances of all the estimators using real EEG data from 89 subjects with both regularization and volume conduction matching problem tested exhaustively.

## 2. Unified reference estimator

In what follows, we denote scalars with lowercase symbols (e.g. $x$), vectors with lowercase bold (e.g. $\mathbf{x}$), matrices with uppercase bold (e.g. $\mathbf{X}$); unknown parameters will be denoted by Greek letters (e.g. $\xi$). Furthermore, $\mathbf{1}$ is the vector of ones; $\mathbf{I}_{N_e}$ is a $N_e$ by $N_e$ identity matrix; $N(\boldsymbol{\mu},\boldsymbol{\Sigma})$ is the multivariable Gaussian distribution with mean vector $\boldsymbol{\mu}$ and covariance matrix $\boldsymbol{\Sigma}$; $(\cdot)^\mathbf{T}$ is the transpose of $(\cdot)$; $\mathbf{X}^+$ is the pseudo-inverse of $\mathbf{X}$; $tr(\cdot)$ is the trace of $(\cdot)$; $\hat{\mathbf{x}}$ is the estimation of $\mathbf{X}$; $\|\cdot\|_\mathbf{F}$ is the Frobenius norm; $\|\cdot\|_M$ is the Mahalanobis distance; $|\cdot|$ is the matrix determinant operator.

### 2.1. General reference model

The EEG is always recorded with respect to a time-varying reference. This is usually modeled as a constant subtracted from all electrodes at each instant. In the general case, we can consider that there are two separate reference constants, one for the scalp EEG signal, and another for the sensor noise (if they come from distinctly different source). In this case, the online recorded EEG signal at a given instant is modeled as

$$\mathbf{v} = \boldsymbol{\varphi} - \mathbf{1} \cdot \rho + \boldsymbol{\varepsilon} - \mathbf{1} \cdot \zeta \tag{1}$$

where $\boldsymbol{\varphi}$ is the pure EEG signal with the neutral reference over $N_e$ electrodes, i.e. abovementioned $\boldsymbol{\varphi}_\infty$, and its distribution is $N(\mathbf{0}, \boldsymbol{\Sigma}_{\varphi\varphi})$; $\boldsymbol{\varepsilon}$ is the sensor noise with $N(\mathbf{0}, \sigma^2 \mathbf{I}_{N_e})$; $\rho$ and $\zeta$ are two reference constants of EEG signal $\boldsymbol{\varphi}$ and sensor noise $\boldsymbol{\varepsilon}$, respectively. $\rho$ is assumed from a cephalic source, but $\zeta$ may come from either cephalic, non-cephalic or the coupled sources. Due to the uncertainty of these constants, the reference of $\mathbf{v}$ is an unknown variable. Note that other distributions for $\boldsymbol{\varphi}$ and $\boldsymbol{\varepsilon}$ may be used with our same general framework.

Applying a reference process is just a linear transformation of EEG data. Formally, it is the pre-multiplication of the reference transformation matrix with the EEG data. Thus, supposing a reference transformation matrix $\mathbf{H} = \mathbf{I} - \mathbf{1}\mathbf{f}^T$ [25], a referential recording is

$$\mathbf{v}_r = \mathbf{H}\mathbf{v} = \mathbf{H}(\boldsymbol{\varphi} + \boldsymbol{\varepsilon}) - (\mathbf{I} - \mathbf{1}\mathbf{f}^T) \cdot \mathbf{1} \cdot (\rho + \zeta)$$

Notably, the equation $\mathbf{f}^T \mathbf{1} = 1$ is satisfied for all the unipolar references, such as monopolar recording references (e.g. Cz, Fz, Oz, etc.), linked mastoids and average reference.

Thus, the general EEG reference model becomes

$$\mathbf{v}_r = \mathbf{H}\boldsymbol{\varphi} + \mathbf{e}, \ \mathbf{e} = \mathbf{H}\boldsymbol{\varepsilon} \qquad (2)$$

where $r$ denotes a specific reference. Note that $\mathbf{H}$ is a rank-1 matrix. Thus, the estimate of $\boldsymbol{\varphi}$ is transformed into an undetermined generalized linear inverse problem.

Taking the penalized maximum log-likelihood estimation (LaRiccia and Eggermont, 2009), we have the objective function

$$l = -\frac{1}{2}\log\left|4\pi^2 \boldsymbol{\Sigma}_{ee} \boldsymbol{\Sigma}_{\varphi\varphi}\right| - \frac{1}{2}(\mathbf{v}_r - \mathbf{H}\boldsymbol{\varphi})^T \boldsymbol{\Sigma}_{ee}^+ (\mathbf{v}_r - \mathbf{H}\boldsymbol{\varphi}) - \frac{1}{2}\boldsymbol{\varphi}^T \boldsymbol{\Sigma}_{\varphi\varphi}^+ \boldsymbol{\varphi} \qquad (3)$$

After finding the partial derivative of (3) with respect to $\boldsymbol{\varphi}$, it follows that

$$\hat{\boldsymbol{\varphi}} = (\mathbf{H}^T \boldsymbol{\Sigma}_{ee}^+ \mathbf{H} + \boldsymbol{\Sigma}_{\varphi\varphi}^+)^+ \mathbf{H}^T \boldsymbol{\Sigma}_{ee}^+ \mathbf{v}_r$$

Referring to the formula (5.137) in page 133 of (Albert Tarantola, 2005), $\hat{\boldsymbol{\varphi}}$ is re-expressed as

$$\hat{\boldsymbol{\varphi}} = \boldsymbol{\Sigma}_{\varphi\varphi} \mathbf{H}^T (\mathbf{H}\boldsymbol{\Sigma}_{\varphi\varphi}\mathbf{H}^T + \boldsymbol{\Sigma}_{ee})^+ \mathbf{v}_r \qquad (4)$$

which is taken as the unified Bayesian estimator in reconstructing EEG potentials at infinity.

To derive the explicit expression of (4), in addition to assuming $\boldsymbol{\Sigma}_{ee} = \sigma^2 \mathbf{H}\mathbf{H}^T$, $\boldsymbol{\Sigma}_{\varphi\varphi}$ is assumed to have one of the following two different forms.

## 2.2. Uncorrelated prior

$$\boldsymbol{\Sigma}_{\varphi\varphi} = \alpha^2 \mathbf{I}_{N_e} \qquad (5)$$

which means that the EEG potentials $\boldsymbol{\varphi}$ have independent priors across all the channels; $\alpha^2$ is the mean of variances of the potentials over each electrode.

Substituting (5) and $\boldsymbol{\Sigma}_{ee} = \sigma^2 \mathbf{H}\mathbf{H}^T$ into (4), it becomes

$$\hat{\boldsymbol{\varphi}} = \mathbf{H}^+ \mathbf{H}\mathbf{v} / (1 + \sigma^2/\alpha^2) \qquad (6)$$

We show that $\mathbf{H}^+\mathbf{H} = \mathbf{I}_{N_e} - \mathbf{1}\mathbf{1}^T/N_e$ which is the average reference transforming matrix in the Appendix. Defining the sensor noise to the scalp EEG signal ratio as $nsr_1 = \sigma^2/\alpha^2$ and $\mathbf{H}_{ar} = \mathbf{I}_{N_e} - \mathbf{1}\mathbf{1}^T/N_e$, (6) is rewritten as

$$\hat{\boldsymbol{\varphi}} = \mathbf{H}_{ar}\mathbf{v}/(1 + nsr_1) \qquad (7)$$

which we shall call the **regularized average reference (rAR)**. It is obvious that the usual AR is the special case of rAR when $nsr_1 = 0$.

## 2.3. Correlated prior

$$\Sigma_{\varphi\varphi} = K_\infty \Sigma_{jj} K_\infty^T \tag{8}$$

which models the EEG potentials across all the channels as correlated due to the effect of volume conduction on neural current sources, that is, we assume that $\varphi = K_\infty j$; $K_\infty$ is the lead field matrix with infinity reference; $j$ is the primal current density of the neural current sources with $j \sim N(0, \beta^2 I_{N_s})$; $N_s$ is the number of neutral current sources; $\beta^2$ is the variance of the multivariate Gaussian signal $j$.

(4) is transformed by substituting (8) and defining $K_r = HK_\infty$ as

$$\hat{\varphi} = K_\infty \cdot \Sigma_{jj} K_r^T (K_r \Sigma_{jj} K_r^T + \Sigma_{ee})^+ v_r \tag{9}$$

which is the estimator for reconstructing the EEG potentials at infinity named as the **regularized reference electrode standardization technique (rREST)**. This process can be interpreted as processing in two stages,

Stage 1: $\hat{j} = \Sigma_{jj} K_r^T (K_r \Sigma_{jj} K_r^T + \Sigma_{ee})^+ v_r$

Stage 2: $\hat{\varphi} = K_\infty \hat{j}$

the first one of which is solving the inverse problem with lead field $K_r$ that has the same reference as the EEG potentials $v_r$ and the second one of which is taking the forward calculation to reconstruct the EEG potentials with the theoretical neutral infinity reference. In stage 1, $\hat{j}$ is the standard form of solving linear inverse problems and the reference problem, simultaneously.

Defining the sensor noise to the brain source signal ratio as $nsr_2 = \sigma^2/\beta^2$ and plugging $\Sigma_{jj} = \beta^2 I_{N_s}$, $\Sigma_{ee} = \sigma^2 HH^T$ into (9), it becomes

$$\hat{\varphi} = K_\infty \cdot K_r^T (K_r K_r^T + nsr_2 \cdot HH^T)^+ v_r \tag{10}$$

which is the solution to reconstruct the EEG potential at infinity through solving the inverse solution by incorporating the identity diagonal structure of $\Sigma_{jj}$. Apparently, REST (Yao, 2001) $\hat{\varphi} = K_\infty \cdot K_r^+ v_r$ is the special case of rREST when $nsr_2 = 0$ in (10).

For clarity, we summarize the general reference model and unified reference estimator in **Table 1.**

**Table 1**, EEG reference model, unified estimator and schemes

| General Reference Model | $v_r = H\varphi + e,\ e = H\varepsilon$ | | | | |
|---|---|---|---|---|---|
| Unified Reference Estimator | $\hat{\varphi} = \Sigma_{\varphi\varphi} H^T (H\Sigma_{\varphi\varphi} H^T + \Sigma_{ee})^+ v_r$ | | | | |
| Prior of $\varphi$ | $\Sigma_{\varphi\varphi} = \alpha^2 I_{N_e}$ | | $\Sigma_{\varphi\varphi} = K_\infty \Sigma_{jj} K_\infty^T$ | | |
| Solutions | $\hat{\varphi} = H_{ar} v/(1+nsr_1)$ | | $\hat{\varphi} = K_\infty \cdot \Sigma_{jj} K_r^T (K_r \Sigma_{jj} K_r^T + \Sigma_{ee})^+ v_r$ | | |
| Prior of $j$ | | | $\Sigma_{jj} = \beta^2 I_{N_s}$ | | $\Sigma_{jj} \neq \beta^2 I_{N_s}$ |
| Sensor noise | zero | nonzero | zero | nonzero | nonzero |
| Reference schemes | AR | rAR | REST | rREST | |

## 3. Reference evaluation

**Table 1** shows that both AR and REST are special cases of rAR and rREST if either the sensor noise is supposed to be zero or no regularization is applied. In this section, after transforming the general reference model into the standard ridge regression form, we evaluate the references via statistical model selection criteria.

### 3.1. Standard regression form

The objective function (3)ion is equivalent to the general ridge regression form

$$\hat{\boldsymbol{\varphi}}(\lambda) = \arg\min_{\boldsymbol{\varphi}}\{\|\mathbf{v}_r - \mathbf{H}\boldsymbol{\varphi}\|_M^2 + \lambda\|\mathbf{L}\boldsymbol{\varphi}\|_2^2\} \qquad (11)$$

where $\lambda \geq 0$ is the regularization parameter; $\mathbf{L}$ is the regularization matrix. For convenience, we call the regularization of $\lambda$ and $\mathbf{L}$ as 'parameter regularization' and 'structure regularization', respectively.

Since $\boldsymbol{\Sigma}_{ee} = \sigma^2 \mathbf{H}\mathbf{H}^T$, if $\mathbf{H} = \mathbf{U}\mathbf{S}\mathbf{V}^T$ and $\mathbf{S}^+ = \mathbf{D}\mathbf{D}^T$, the noise term $\mathbf{e}$ is redefined as $\mathbf{e}' = \mathbf{D}^T\mathbf{U}^T\mathbf{e}$. Thus, the general ridge regression form (11)into the standard ridge regression form as to the standard ridge regression form as

$$\boldsymbol{\varphi}(\lambda) = \arg\min_{\boldsymbol{\varphi}}\{\|\mathbf{v}_r' - \mathbf{H}'\boldsymbol{\varphi}'\|_2^2 + \lambda\|\boldsymbol{\varphi}'\|_2^2\} \qquad (12)$$

by redefining $\mathbf{v}_r' = \mathbf{D}^T\mathbf{U}^T\mathbf{v}_r$, $\mathbf{H}' = \mathbf{D}^T\mathbf{U}^T\mathbf{H}\mathbf{L}^+$ and $\boldsymbol{\varphi}' = \mathbf{L}\boldsymbol{\varphi}$. Then, the posterior mean of $\boldsymbol{\varphi}'$ given $\mathbf{v}_r'$ is

$$\hat{\boldsymbol{\varphi}}' = (\mathbf{H}'^T\mathbf{H}' + \lambda\mathbf{I}_{N_e})^+ \mathbf{H}'^T\mathbf{v}_r' \qquad (13)$$

then, the estimate of $\boldsymbol{\varphi}$ is $\hat{\boldsymbol{\varphi}} = \mathbf{L}^+(\mathbf{H}'^T\mathbf{H}' + \lambda\mathbf{I}_{N_e})^+ \mathbf{H}'^T\mathbf{v}_r'$ which is equivalent to the formula (10).

### 3.2. Model selection criteria

Since ridge regression is a linear estimator ($\hat{\mathbf{v}}_r' = \mathbf{P}\mathbf{v}_r'$) with $\mathbf{P} = \mathbf{H}'(\mathbf{H}'^T\mathbf{H}' + \lambda\mathbf{I}_{N_e})^+ \mathbf{H}'^T$ where $\mathbf{P}$ is the projection ('hat') matrix. The residual sum square error (RSS) is defined as

$$\text{RSS} = \sum_{t=1}^{N_t} \|\mathbf{v}_{rt}' - \mathbf{H}'\hat{\boldsymbol{\varphi}}_t'\|_2^2$$

where $\mathbf{v}_{rt}'$ and $\hat{\boldsymbol{\varphi}}_t'$ with subscript $t$ denote $\mathbf{v}_r'$ and $\hat{\boldsymbol{\varphi}}'$ at the $t^{th}$ ($t = 1, \cdots, N_t$) time sample, respectively; $N_t$ is the number of time samples in the whole EEG recording.

Under the standard ridge regression form (12)explore three information criteria for the model selection: generalized cross-validation (GCV), Akaike information criteria (AIC), and Bayesian information criteria (BIC) to compare the reference schemes in **Table 1**. To apply these, we define the degree of freedom (DF) as the degree of freedom (DF) as

$$\text{DF}(\lambda) = tr(\mathbf{P}) = \sum_{i=1}^{N_e} \frac{s_i}{s_i + \lambda}$$

where $\{s_i\}$ are the eigenvalues of $\mathbf{H}'^T\mathbf{H}'$. Since EEG reference acts as adding or subtracting a time-varying constant over all sensors at each instant, this instantaneous effect results in the dynamical alteration in the temporal domain. To investigate the difference of references, we extend the model selection criteria from single instant to the whole recording, approximately. Predefining $N_{et} = N_e \cdot N_t$, GCV, AIC and BIC are expressed as

$$\text{GCV}(\lambda) = \text{RSS}/(N_{et} - \text{DF})^2 \tag{14}$$

$$\text{AIC}(\lambda) = N_{et} \log(\text{RSS}/N_{et}) + N_t \cdot 2 \cdot \text{DF} \tag{15}$$

$$\text{BIC}(\lambda) = N_{et} \log(\text{RSS}/N_{et}) + N_t \cdot \text{DF} \cdot \log(N_{et}) \tag{16}$$

Note that GCV, AIC, and BIC at a single instant are the special cases of (14)–(16), respectively $N_t = 1$, respectively.

### 3.3. Regularization parameter

The regularization parameter $\lambda$ balances the goodness of fitting (i.e. likelihood) and the prior constraint on the EEG potentials at infinity. One may try to interactively estimate the hierarchal Bayesian hyperparameter via iteration (MacKay, 1992; Trujillo-Barreto et al., 2004). However, this may work for rREST but poorly for rAR because the noise term will be assimilated into the pure EEG signal in (6) the uncorrelated covariance prior. Namely, the objective function of AR is nonconvex and it cannot converge to a global or local optimal point. Thus, we adopt a search strategy to explore how DF, GCV, AIC and BIC vary with the values of $\lambda$. The idea is to plot the DF against $\lambda$ as well as GCV, AIC and BIC against DF. In theory, the optimal $\lambda$ is around $nsr_1$ for rAR, and approximates to $nsr_2$ for rREST, respectively. Since volume conduction acts as a lowpass spatiotemporal filter resulting in $nsr_2 \ll nsr_1$, we thus only generate 1000 values of $\lambda$ from 1e-3.5 to 1e-1 for rREST, and 1e-3 to 10 for rAR, by using sampled logarithm.

In the simulation, we can evaluate the reference estimators with an 'oracle' regularization parameter, namely, one for which the smallest relative error regarding the ground truth. Additionally, the efficiency of the model selection criteria (GCV, AIC and BIC) for selecting the regularization parameter is evaluated. It will be trickier to find a proper $\lambda$ with actual EEG data where the ground truth is unknown. The value with which one model selection criteria reaches to a global or local minimum is regarded as the optimal $\lambda$ chosen by the model selection criteria for actual EEG data.

It has been suggested to avoid regularization when applying REST so as not to lose high-frequency information (Yao, 2001). Instead, a truncation of singular value decomposition (TSVD) was proposed to suppress the effect of sensor noise for REST (Zhai and Yao, 2004). Therefore, we empirically adopt the recommended truncation parameter 0.05 for REST but use the model selection criteria for rREST.

### 3.4. Regularization matrix

The choice of regularization matrix $\mathbf{L}$ depends on the prior covariance structure of the potentials at infinity. **Table 1** shows that the prior covariance structure of $\boldsymbol{\varphi}$ as $\boldsymbol{\Sigma}_{\boldsymbol{\varphi\varphi}} = \alpha^2 \mathbf{I}_{N_e}$ for AR and rAR, and $\boldsymbol{\Sigma}_{\boldsymbol{\varphi\varphi}} = \mathbf{K}_{\infty} \boldsymbol{\Sigma}_{\mathbf{jj}} \mathbf{K}_{\infty}^{\mathbf{T}}$, for REST and rREST, respectively. Therefore, the choices of $\mathbf{L}$ are:

for AR and rAR, $\qquad \mathbf{L}_{ar} = \mathbf{I}_{N_e}$

for REST and rREST, $\qquad \mathbf{L}_{rt} = [(\mathbf{K}_{\infty} \mathbf{K}_{\infty}^{\mathbf{T}})^{+}]^{1/2}$ (17)

Several cases of $\mathbf{K}_{\infty}$ are detailed in the next section. The degree of faithful biophysical regularization by $\mathbf{L}_{rt}$ increases from the less realistic approximation of volume conductor to the more realistic one.

### 3.5. Volume conduction model

For rREST, we study the volume conduction model matching problem, that is, to what extent, the lead field for rREST may be different from the actual one that generated the simulated EEG data or the real EEG recordings. Here, we evaluate several types of lead fields. The well-known spherical lead field (SLF) is a frequently adopted standard lead field. The most precise volume conduction model is the individual lead field (ILF) matched to the structural Magnetic Resonance Image (sMRI) of each subject. We also evaluate the average lead field (ALF) as a substitute for the individual lead field. Finally, we evaluate the 'sparse individual lead field' (sILF) for which we switch off the voxels not used in the simulations. We will use suffixes to distinguish between types of lead fields.

**Spherical lead field (SLF)** $\mathbf{K}_\infty^s$, is estimated based on the standard 3-layers concentric spherical head model comprising of brain, skull, and scalp with the conductivities being 1, 0.0125 and 1, respectively. For the spherical head shape, the radii are 1.0, 0.92 and 0.87 for the scalp surface, outer and inner skull surface, separately. The source space consists of 2600 discrete dipole sources evenly and radially distributed on the cortical surface with radius r = 0.86 and 400 discrete dipole sources uniformly located perpendicularly to the transverse plane with Z = - 0.076 (for the coordinates in detail, refer to (Hu et al., 2017; Yao, 2001)).

**Individual lead field (ILF)** $\mathbf{K}_\infty^i$, is defined by normalization as

$$\mathbf{K}_\infty^i = \mathbf{K}_\infty^{iraw} \Big/ [tr(\mathbf{K}_\infty^{iraw} \mathbf{K}_\infty^{iraw\mathbf{T}})]^{1/2}$$

where $\mathbf{K}_\infty^{iraw}$ is the raw individual lead field matched to the $i^{th} (i=1,\cdots,N_b)$ subject who underwent the EEG recording in Cuban Human Brain Mapping Project (CHBMP) (Bosch-Bayard et al., 2012; Hernandez-Gonzalez et al., 2011; Uludağ et al., 2009; Valdés-Hernández et al., 2010). It is estimated by the finite element method based on the segmented cortical surface through CIVET pipeline (Ad-Dab'bagh, Y., Lyttelton, O., Muehlboeck, J. S., Lepage, C., Einarson, D., Mok, K., ... & Evans, 2006) with sMRI. The cortical surface is formed by 6003 vertices and 11998*3 faces. In total, 6003 dipole sources are located at the vertices and activated perpendicularly to the cortical surface, individually. The normalization allows for comparison across subjects.

**Average lead field (ALF)** $\mathbf{K}_\infty^a$, is the average of all the normalized ILFs of $N_b$ subjects as

$$\mathbf{K}_\infty^a = \frac{1}{N_b} \sum_{\mathbf{i}=1}^{N_b} \mathbf{K}_\infty^i$$

**Sparse individual lead field (sILF)** $\mathbf{K}_\infty^{si}$, (for use in simulation) is obtained after transforming ILF as follows

$$\mathbf{K}_\infty^{si} = [\mathbf{K}_\infty^{iraw} \circ \mathbf{W_i}] \Big/ [tr(\mathbf{K}_\infty^{iraw} \circ \mathbf{W_i} \mathbf{W_i^T} \circ \mathbf{K}_\infty^{iraw\mathbf{T}})]^{1/2}$$

where $\circ$ means the matrix elementwise multiplication operation (i.e. Hadamard product); $\mathbf{W_i}$ is a matrix that consists of binary weights and has the same size with $\mathbf{K}_\infty^{iraw}$; the entries at the columns of un-activated brain sources are zeros and the other columns are full of ones. In the simulation, the position of two patches of sources is incorporated into the covariance of the EEG potentials at infinity for rREST. In place of adopting $l_0$ norm or $l_1$ norm to sparse the brain electrical source signal $\mathbf{j}$, we set the entries corresponded to non-activated sources of ILF being zeros to constrain the brain source signal indirectly.

# 4. Simulation

## 4.1. EEG generation

The simulation scheme is based on the forward equation below,

$$\begin{cases} \mathbf{v}_r = \mathbf{H}\boldsymbol{\varphi} + \mathbf{H}\boldsymbol{\varepsilon}, \boldsymbol{\varphi} = \mathbf{K}_\infty^{iraw} \mathbf{j} \\ \text{SNR} = 10\log_{10}(\alpha^2/\sigma^2) \end{cases} \quad (18)$$

where $\mathbf{v}_r$ is the simulated EEG potentials with unipolar reference; without loss of generality, the linear combination vector $\mathbf{f}=[0,\cdots,0,1]^\mathrm{T}$ with the last entry being one and the others being zeros; Two patches consisting of 150 dipole sources in $\mathbf{j}$ are activated, meeting 4-order bivariate autoregressive model (**Figure 3**); SNR is the scalp EEG signal to the sensor noise variance ratio in dB unit.

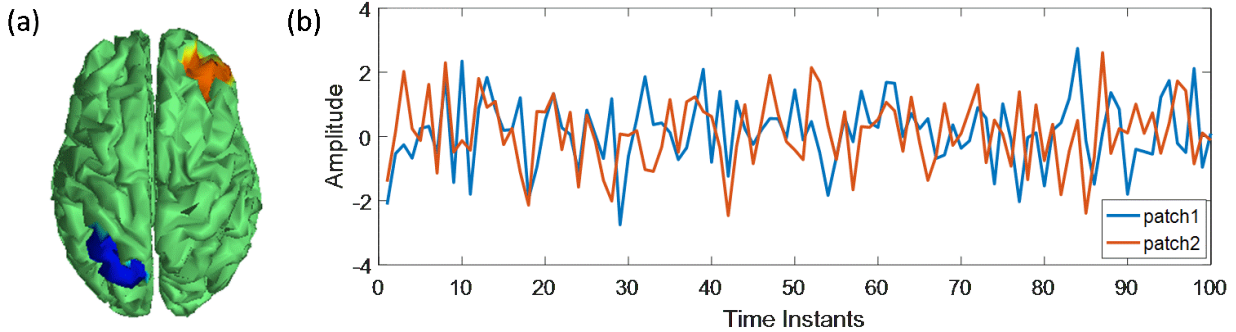

**Figure 3, Source activities for EEG simulation. (a)** location of two source patches, **(b)** activities of two source patches.

With one $\mathbf{K}_\infty^{\text{iraw}}$, the simulated EEG are 58 channels by 5120 instants. Individually, taking the $\mathbf{K}_\infty^{\text{iraw}}$ of 89 subjects from the CHBMP database, we generated the EEG dataset 89 samples by 58 channels by 5120 instants. The simulation provides the ground truth of EEG potentials with the neutral reference, thus making it possible to intuitively compare the performances of references in terms of the relative error of potentials. The relative error (RE) of potentials is defined as

$$\mathrm{RE} = \|\hat{\boldsymbol{\varphi}} - \boldsymbol{\varphi}\|_\mathrm{F}^2 / \|\boldsymbol{\varphi}\|_\mathrm{F}^2 \tag{19}$$

where $\varphi$ denotes the ground truth; $\hat{\varphi}$ is the EEG potentials estimated by the references in **Table 1**.

### 4.2. Relative error of reference estimators

For each data sample (58 channels by 5120 instants), the relative error (RE) of potentials is calculated by using (19). **Figure 4(a)-(d)** show the REs of the reference estimators, including the lead fields variants (SLF, ILF, ALF, and sILF) for REST and rREST, with SNR=20dB, 8dB, 4dB, 2dB, respectively. Boxplots in black, green, red, and blue, show the REs of AR, rAR, REST and rREST, separately. It is evident from the boxplots **(a)-(d)**, that the REs of regularized references (rAR, rREST) are always less than that of unregularized references (AR, REST). Unpaired t-tests were applied to check the differences between unregularized references (AR, REST) and regularized references (rAR, rREST). **Figure 4(e)** lists the statistical significance levels (p values) between AR and rAR, as well as between REST and rREST with various lead fields tested, separately. Except for the case between AR and rAR with SNR=20dB, the p-values all reach very small values (<1e-7).

With regularization, the decreases of REs from REST to rREST are more obvious than the decreases of REs from AR to rAR. Especially, regularization with sILF is much more efficient than SLF, ILF, and ALF. This is not surprising since the sparse prior information was incorporated into the covariance structure. By contrast, by the simplest volume conduction model, i.e. SLF, the REs of rREST seems to be even larger than that of AR, and REST performs the worst among all the references, when SNR = 20dB and 8dB. Comparing the REs by sILF and that of rREST by SLF with the REs of REST by SLF, we found that structure regularization by precise covariance seems to be more efficient than the parameter regularization by selecting the optimal $\lambda$. And the REs of rREST with sILF are the least among all the REs of other references. This means that structure regularization together with parameter regularization will have the best effect. In addition, injecting higher sensor noise with SNR being from 20dB to 2dB, the REs of rAR increase from less than 15% to higher than 60% accordingly, while the REs of rREST with SLF excluded goes from 4.1% to 40%.

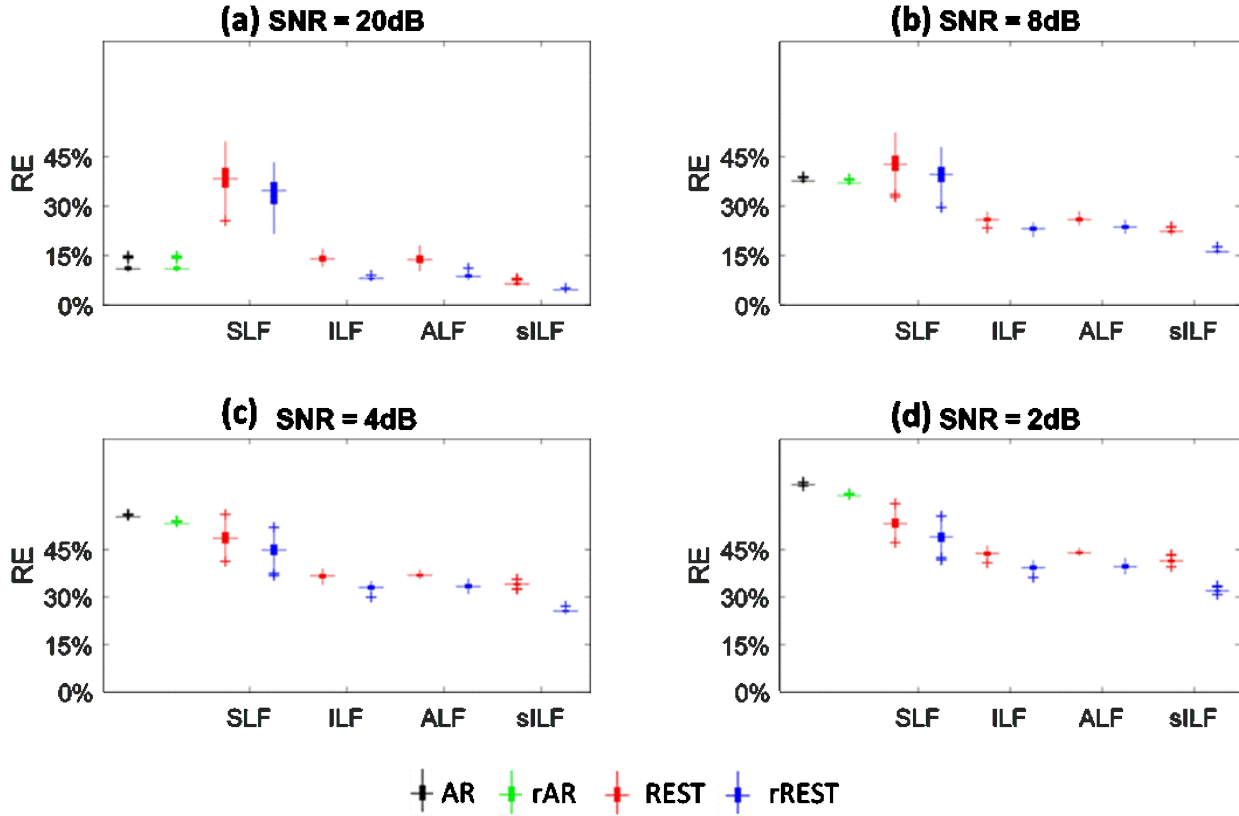

**Figure 4, Relative error (RE) of reference estimators. (a) – (d),** the boxplots of REs with SNR = 20dB, 8dB, 4dB and 2dB, respectively. Volume conduction model tested for REST and rREST are spherical lead field (SLF), individual lead field (ILF), average lead field (ALF), and sparse individual lead field (sILF). **(e)** the p-values of REs between ordinary references (AR, REST) and regularized references (rAR, rREST) under different SNR and various lead fields, separately.

These results indicate that: 1) except for the case of AR and rAR with SNR=20dB, AR, rAR, REST and rREST by using SLF that roughly approximated the actual volume conduction model may be not able to reconstruct the EEG signal at infinity due to the quite large REs; 2) the effects of REST and rREST are volume conduction model dependent; 3) Stronger regularization applied, better effect of rREST obtained; 4) for REST and rREST, the REs by using ALF seems to be almost same with the REs by ILF; 5) rAR may not have the effect of denoising. Over all, AR and rAR may be the alternative option when SNR is very large (>=20dB), while rREST with precise volume conduction model should be the first option to estimate the EEG signal at infinity.

## 4.3. Model selection for estimators with simulated data

The results summarized in **Figure 5** allow determining the optimal reference via the model selection criteria. The curves in **(a)** show how the degree of freedom (DF) and Generalized Cross Validations (GCV) vary with the regularization parameter $\lambda$ (i.e. LMD), as well as how the residual sum square (RSS) changes with DF. It is easy to see that the DF of rREST are always smaller than the DF of rAR. This means that rREST adopts the simpler model to reconstruct the EEG signal at infinity but employ the more realistic prior information for regularization than rAR. The lower RSS of rREST than rAR indicate that the EEG signal reconstructed by rREST is closer to the truth compared with the EEG signal restored by rAR. The curves in **(b)** display how the model selection criteria, GCV, AIC and BIC vary with the DF. Apparently, the model selection criteria values of rREST are always smaller than them of rAR. The prevalent lower values of model selection criteria provide the evidence to prefer rREST over rAR.

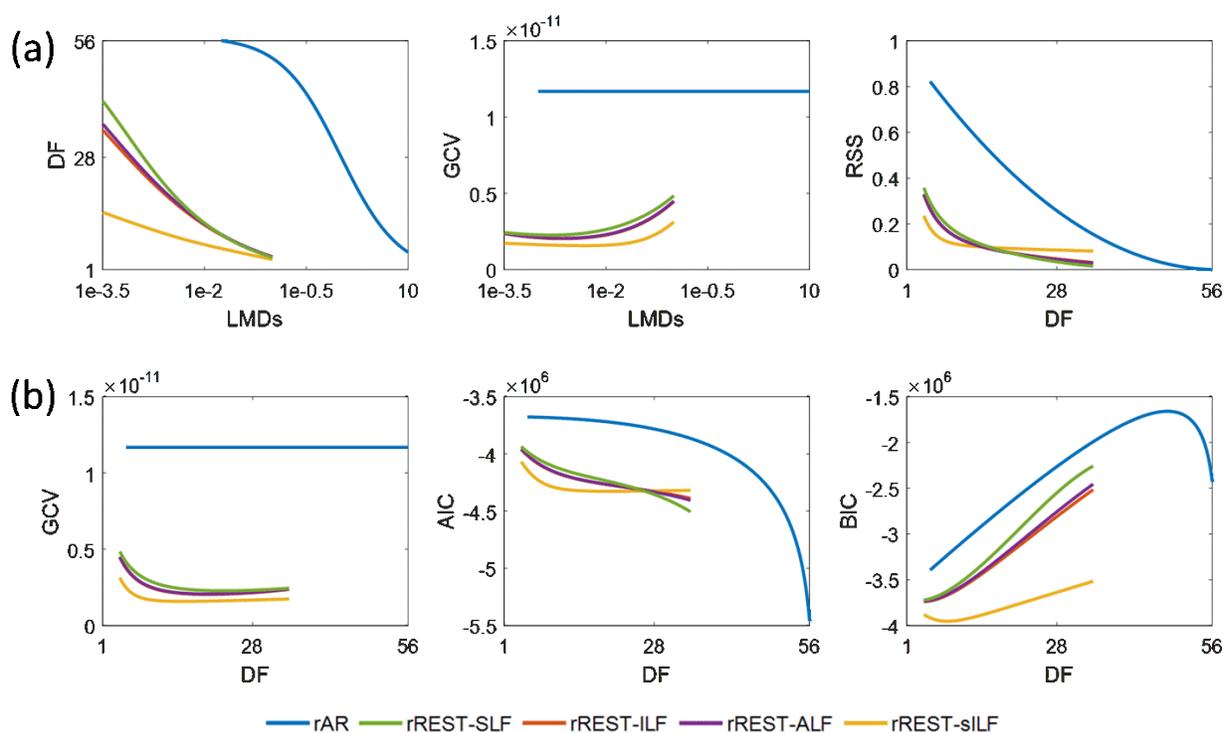

**Figure 5, Model selection with simulated data.** (a), DF and GCV against LMD, and RSS varying with DF; (b), model selection criteria (GCV, AIC and BIC) against DF. DF ---- degree of freedom, RSS ---- residual sum square, GCV ---- generalized cross validation, AIC ---- Akaike information criteria, BIC ---- Bayesian information criteria, SLF ---- the lead field with 3-layers spherical head model, ILF ---- the normalized individual lead field built with realistic head model, ALF ---- the average of normalized individual lead fields built with realistic head model over 89 subjects, sILF ---- the normalized individual lead field with sparse prior information.

## 4.4. Regularization parameter

For rREST, it is crucial to pick the best regularization parameter, i.e. the value of $\lambda$. **Figure 6** displays that, to what extent, the values of $\lambda$ selected by the model selection criteria (GCV, AIC and BIC) are close to the truth. Comparing the mean relative error (mRE) and the standard deviations in **(b)**-**(d)** with those in **(a)**, GCV is easily found as the best one to select the proper regularization parameter due to the almost same mRE and standard deviations to the truth; AIC is worse than GCV since except for the rREST by using sILF, the regularized reference (rAR, rREST) show the same or larger mRE and standard deviations than the ordinary reference (AR, REST); BIC is the worst one to select the proper regularization parameters because all the regularized references (rAR, rREST) present the larger mREs and standard deviations than the ordinary references (AR, REST).

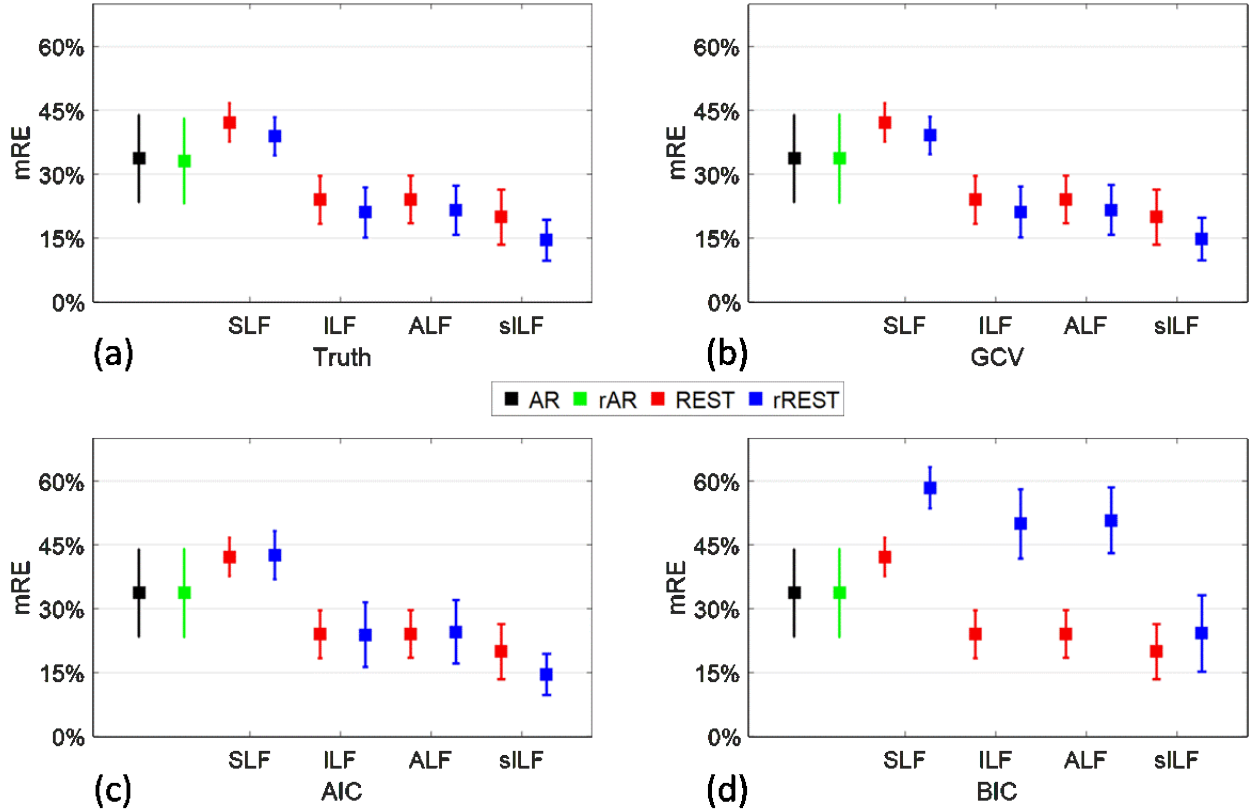

**Figure 6, Regularization parameter selection.** Each square and the error bar are the mean relative error (mRE) and standard deviation over 89 simulations in each of which different raw ILF were used. (a), the truth where the best $\lambda$ is picked by the least RE; (b), (c) and (d), the results where the best $\lambda$ selected by the least GCV, AIC and BIC values, respectively.

## 5. Model selection for estimators with real data

We take the EEG of 89 subjects from the CHBMP database to evaluate the reference estimators. The EEG recordings were carried out in accordance with the recommendations of Ethics committees of Ministry of Public health and Cuban Neuroscience Center with written informed consent from all subjects. The EEG was acquired with 58 channels, 10-10 electrode placement system, sampling rate 200Hz, recording period 2.5-5 mins, and with the resting-state of 'eyes-closed-open' intermittently conditioned. **Figure 7** displays a real EEG data sample. To compare the performance of references over all subjects, we normalize the EEG data by

$$\mathbf{v}_r = \mathbf{v}_r^{raw} \big/ \left\| \mathbf{v}_r^{raw} \right\|_F$$

The model selection criteria GCV, AIC, and BIC are calculated for each subject with the matched ILF, sILF, the identical ALF and SLF. The mean model selection criteria are averaged over 89 subjects.

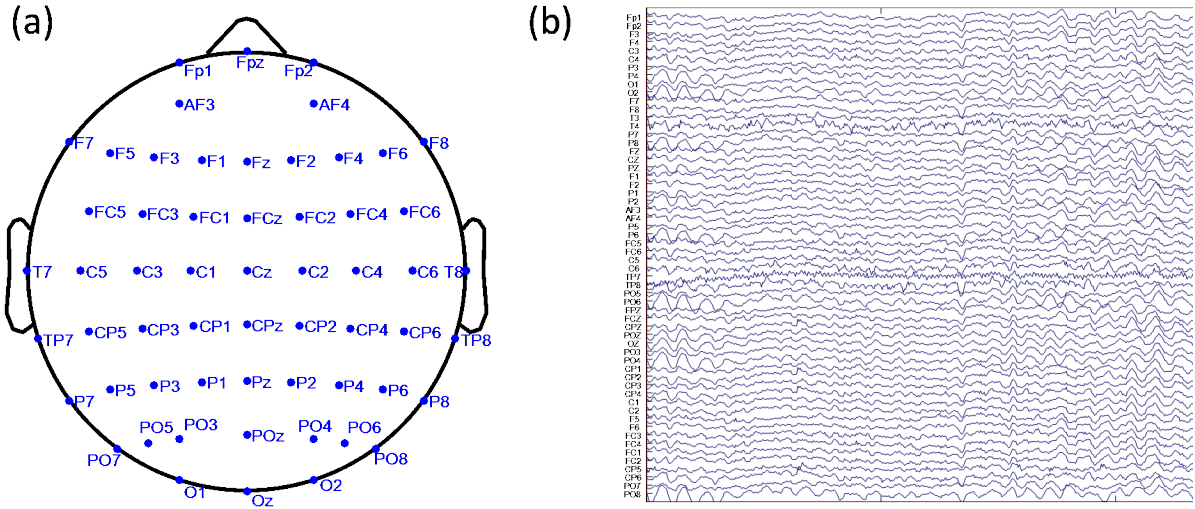

**Figure 7, The illustration of a real EEG data sample. (a)** the electrodes position, **(b)** the waveform of one epoch.

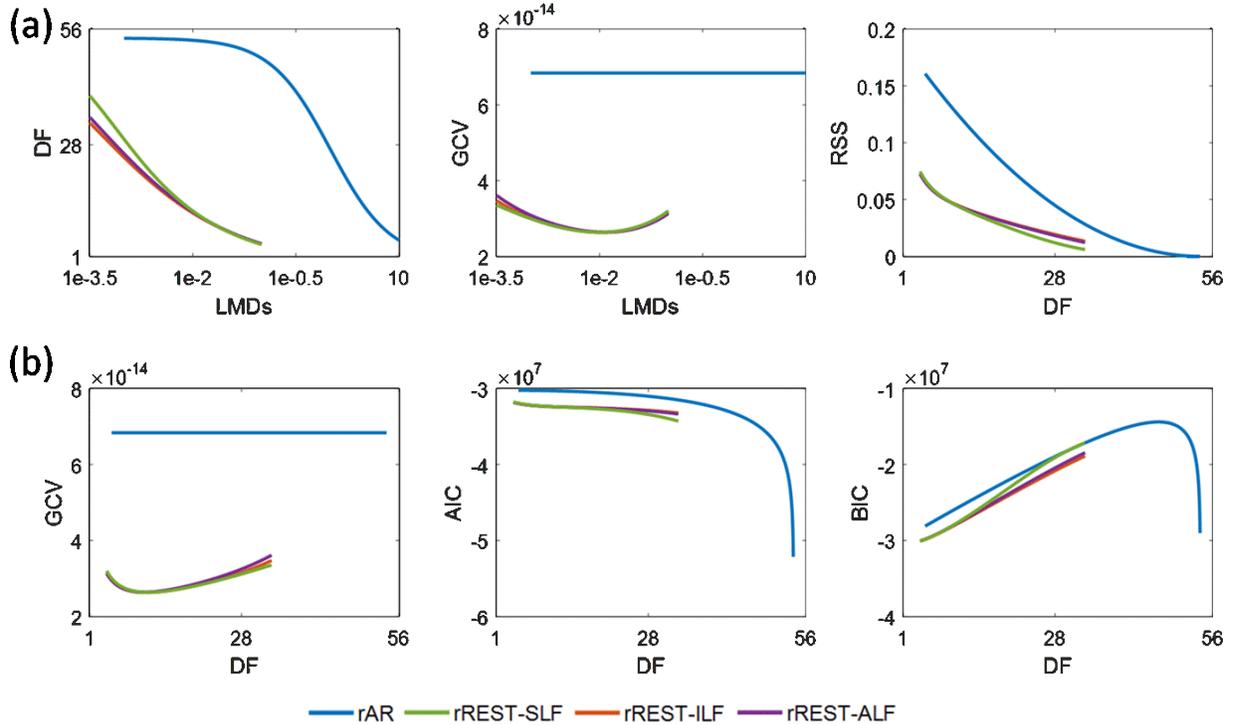

**Figure 8**, **Model selection with actual EEG**. (a), DF and GCV against LMD, and RSS varying with DF; (b), model selection criteria (GCV, AIC *and* BIC) against DF. DF ---- degree of freedom, RSS ---- residual sum square, GCV ---- generalized *cross validation*, AIC ---- Akaike information criteria, BIC ---- Bayesian information criteria, SLF ---- the lead field with 3-layers spherical head model, ILF ---- the normalized individual lead field built with realistic head model, ALF ---- the average of normalized individual lead fields built with realistic head model over 89 subjects, sILF ---- the normalized individual lead field with sparse prior information.

To validate the preference of rREST over rAR, a similar analysis as descried before for simulations regarding performance is now applied to the realistic resting state EEG datasets of 89 subjects from the CHBMP database. The results in **Figure 8** show how the degree of freedom (DF) and generalized cross validation (GCV) change with the regularization parameter $\lambda$ (i.e. LMD) and how the residual sum square (RSS) and model selection criteria (GCV, AIC, and BIC) vary with the DF. Since the reference transformation matrix $\mathbf{H}$ and the limits of regularization parameter $\lambda$ (i.e. LMD) are the same as that in simulation, the curves of DF against LMDs in **Figure 8 (a)** are identical to them in **Figure 5 (a)**. The lower RSS and model selection criteria curves in **Figure 8 (b)** confirm, for real data, our previous findings in simulations, namely, that 1) the EEG signal reconstructed by rREST has lower RSS than that restored by rAR; 2) rREST has the smaller values of GCV, AIC, and BIC than rAR, except for the almost same BIC between rREST and rAR around DF = 28. Due to that GCV is found to be the best to pick the proper regularization parameter $\lambda$ in **Figure 6**, we suggest adopting GCV to select the value of $\lambda$ in practice when the ground truth is unknown. The curves in the middle of **Figure 8 (a)** shows how GCV varies with the values of $\lambda$ (i.e. LMDs). For rREST, the global minimum of GCV occurs around $\lambda$ = 1e-2 or about DF = 10. This means that for a group of subjects, we are able to find the best regularization parameter by the lowest GCV. By contrast, GCV of rAR seems to be one constant. In other words, the solution of rAR is a nonconvex solution where it is hard to find the proper $\lambda$. These results indicate that the preference of rREST over rAR is validated for real EEG. Moreover, for rREST. the best regularization parameter can be picked at the global minimum of GCV curves.

# 6. Discussion

Although the reference electrode standardization technique (REST) was put forward some time ago, its theoretical underpinnings have not been deeply studied, particularly from a mathematical statistics perspective. Prior to REST, the average reference (AR) had been broadly adopted, e.g. in microstates analysis (Khanna et al., 2015). Currently, both AR and REST are the main competing estimators (Nunez, 2010b). Many comparative studies have been carried out, however, without providing the definitive evidence to prefer one over another (Chella et al., 2016, 2017; Lei and Liao, 2017; Qin et al., 2010). The need to settle this issue has been reinforced recently by the theoretical results of Yao, who demonstrated that the main assumption of AR — the cancellation of brain potentials averaged over the scalp is, in general, false (Yao, 2017). However, it is difficult to decide the reference issue solely the biophysical interpretations. Empirical verification of the best reference using a full statistical model is also essential.

In this study, we propose to view the estimation of the potentials at infinity and the determination of reference as a linear inverse problem that can be attacked using well known Bayesian techniques. To our surprise, both AR and REST are two special cases with different prior distributions for the covariance of the EEG potentials referenced to infinity. By explicitly introduced the sensor noise term into the reference model, we combined the estimation of the potentials at infinity with denoising. The formulation, based on penalized maximum likelihood, leads to the regularized estimators, rAR and rREST. Finally, recognizing that the reference is a linear inverse problem leads to the use of model selection criteria to examine several issues. It is found, for both simulated and actual data, that 1) regularization is critical to solving the reference problem and denoising simultaneously; 2) the regularized reference (rAR/ rREST) has a better performance than the ordinary reference (AR/REST), respectively; 3) rREST outperforms REST, 4) to apply rREST to real EEG data, generalized cross-validation is recommended as an effective measure to select the optimal regularization parameter. In our opinion, the definitive argument in favor of rREST is that for 89 resting state EEGs it provides a smaller Generalized Cross Validation. This is the first empirical comparison of references using an information criterion that approximates the Bayesian model evidence.

REST has attracted attention due to its theoretically sound basis (Kayser and Tenke, 2010; Nunez, 2010a; Yao, 2001). However, some studies with ordinary REST suggest that it does not uniformly dominate AR (Hu et al., 2017; Yao, 2001). Though ordinary REST was found to be more efficient than AR for vertically oriented and shallow dipole sources, this was not so for transverse or deep dipoles. Our results give a different perspective. We suggest these findings were due to two factors which are from both a spherical volume conductor model, as well as limiting sources to the equivalent distributed dipoles layer (Yao and He, 2003), i.e. the sources over the 2D cortical sheet with radial orientation (Yao, 2001), to derive the covariance structure for ordinary REST. By contrast, we tested here more realistic volume conduction models. Also, our simulations were based on multiple cortical patches. Our results indicate that more realistic the volume conductor and source space is, the better the estimator of the reference, and that in fact, REST does dominate AR in all cases.

We emphasize multi-possibilities of source modeling and restate the conception of the generalized inverse problem. As REST is a generalized inverse solution, the multipole sources (Daunizeau et al., 2006) and the general 3D distributed sources at each grid of brain volume (Michel et al., 2004) can be also be adopted for REST as well (Yao, 2000). We have shown that all generalized inverses are not equal and an interesting line of research will be to explore how different source model procedures can be translated into variants of REST.

The importance of an adequate volume conduction model matching test showed that REST and rREST is volume conduction model dependent. This is in agreement with the findings of Hu, et al (Hu et al., 2017) and Quanying, et al (Liu et al., 2015). Quanying, et al reported that a realistic volume conduction model is critical to ordinary REST. Hu, et al stated that ordinary REST is volume conduction model dependent but imprecise or slightly perturbated lead fields does not deteriorate ordinary REST much. This is agreement with our simulation results that showed that better estimates of both the volume conductor and source extent lead to better reference estimates.

However, it is computational costly to estimate the individual lead field and is essential to have the subject's sMRI—something not usual in many clinical settings. The excellent performance of the average lead field almost achieves the same performance as obtained with the more exact individual lead field. This validates the proposal that approximate head models without individual MRI data can be quite useful (Valdés-Hernández et al., 2009).

Main contributions of this paper are:

**1**) We propose that reference estimation is a unified Bayesian linear inverse problem.

**2**) This framework explicitly models sensor noise as a part of the EEG generative model. It allows denoising together with reference estimation.

**3**) AR and REST are shown to be the special cases of the linear inverse problem, with a spatially uncorrelated prior for AR and a spatially correlated prior for REST (resulting from volume conduction of sources).

**4**) Regularized estimators, rAR and rREST, are developed to implement reference estimation and denoising simultaneously.

**5**) We adopted the model selection criteria (GCV, AIC, BIC) for not only to select the hyperparameter but also to compare model families. GCV was found to be the most useful indicator.

**6**) We propose the average lead field as a practical substitute for the individual lead field to construct near optimal estimators.

Several extensions of this study are being explored:

- Artifact suppression may be incorporated together with reference estimation and denoising. For example, outliers can be eliminated by utilizing a likelihood function designed for robust statistics (Huber and Ronchetti, 2009).

- More biophysical information may be built into the prior distributions to more effectively differentiate the pure EEG signal from the sensor noise. Particularly, covariance matrices corresponding to different types of structured sparsity source models should be examined (Paz-Linares et al., 2017).
- We have dealt only with spatial priors for the covariance matrix of the EEG. Dynamical priors can easily be incorporated. Temporal autocorrelations may be modeled as state space models and estimated via the Kalman filter (Galka et al., 2004). Alternatively, formulations for the frequency, or time frequency domain are possible.
- The framework may be also extended to event related potentials incorporating prior work from our group in this direction (Carbonell et al., 2004).

# 7. Conclusion

We state the EEG reference problem as a unified inverse problem that can be solved via Bayesian techniques. To our best knowledge, this is a novel approach to the problem. This formulation allows us to:

- Adopt regularization methods to estimate the potential referenced to infinity.
- Demonstrate that REST and AR can be are the special cases of the unified estimator with different EEG spatial covariance priors
- Simultaneously carry out denoising as part of the estimation procedure
- Use model selection criteria to determine the optimal reference estimator. These results can be summarized as:
  - Regularized references (rREST or rAR) are superior to the ordinary REST or AR, with rREST having the overall best performance for both simulations and real data.
  - The optimal choice of volume conductor model is the individual or averaged lead field.

Regularized REST (rREST) may be used in clinical settings, as an improved estimator of EEG potentials referenced to infinity.

## Appendix: Demonstration that $\mathbf{H}^+\mathbf{H} = \mathbf{I} - \mathbf{1}\mathbf{1}^\mathbf{T}/N_e$

We now demonstrate the relation $\mathbf{H}^+\mathbf{H} = \mathbf{I} - \mathbf{1}\mathbf{1}^\mathbf{T}/N_e$. Recall $\mathbf{f}^\mathbf{T}\mathbf{1} = 1$, $\mathbf{H} = \mathbf{I} - \mathbf{1}\mathbf{f}^\mathbf{T}$ which reduces the matrix rank by one. Keeping same form with the formula (1.2) in (Baksalary et al., 2003), we rewrite

$$\mathbf{M} = \mathbf{A} + \mathbf{b}\mathbf{c}^\mathbf{T}, \text{ with } \mathbf{M} = \mathbf{H}, \mathbf{A} = \mathbf{I}, \mathbf{b} = -\mathbf{1}, \mathbf{c} = \mathbf{f}$$

Referring to the **Theorem 1.1** in (Baksalary et al., 2003), it follows that $rank(\mathbf{M}) = rank(\mathbf{A}) - 1$, since $\lambda = 1 + \mathbf{c}^\mathbf{T}\mathbf{A}^+\mathbf{b} = 1 + \mathbf{f}^\mathbf{T}(-1) = 0$ and both $\mathbf{b}$ and $\mathbf{c}$ belong to the column space of $\mathbf{A} = \mathbf{I}$. By utilizing the case (↓) **List 2** in **Theorem 2.1** in (Baksalary et al., 2003), we take

$$\mathbf{M}^+\mathbf{M} = \mathbf{A}^+ - \delta^{-1}\mathbf{d}\mathbf{d}^\mathbf{T}$$

The formulas in (1.3) and (1.4) from (Baksalary et al., 2003) are $\mathbf{d} = \mathbf{A}^\mathbf{T}\mathbf{b}$ and $\delta = \mathbf{d}^\mathbf{T}\mathbf{d}$. Applying these relations to our problem, it turns to be

$$\mathbf{H}^+\mathbf{H} = \mathbf{I} - \mathbf{1}\mathbf{1}^\mathbf{T}/N_e$$

which is the classical average reference.

## Acknowledgements

The authors declare no conflict of interest. This research was co-funded by the National Natural Science Foundation of China projects (NSFC Grants No. 61673090 and 81330032). We thank Esin Karahan and Pedro A. Valdes-Hernandez for their valuable technical assistance.